\documentclass[sigconf]{acmart}
\usepackage{graphicx} 
\usepackage{subfig}
\usepackage{algpseudocode}
\usepackage[ruled,vlined,linesnumbered]{algorithm2e}
\usepackage{breqn}
\usepackage{amsmath}

\usepackage[english]{babel}
\usepackage{amsthm}
\usepackage{enumitem}

\title[Local Differential Privacy for Number of Paths and Katz Centrality]{Local Differential Privacy for \\ Number of Paths and Katz Centrality}
\author{Louis Betzer}
\affiliation{
  \institution{Ecole Polytechnique}
  \city{Palaiseau}
  \country{France}
}
\email{louis.betzer@polytechnique.edu}

\author{Vorapong Suppakitpaisarn}
\affiliation{
  \institution{The University of Tokyo}
  \city{Tokyo}
  \country{Japan}
}
\email{vorapong@is.s.u-tokyo.ac.jp}

\author{Quentin Hillebrand}
\affiliation{
  \institution{The University of Tokyo}
  \city{Tokyo}
  \country{Japan}
}
\email{quentin-hillebrand@g.ecc.u-tokyo.ac.jp}

\begin{abstract}
In this paper, we give an algorithm to publish the number of paths and Katz centrality under the  local differential privacy (LDP), providing a thorough theoretical analysis. Although various works have already introduced subgraph counting algorithms under LDP, they have primarily concentrated on subgraphs of up to five nodes. The challenge in extending this to larger subgraphs is the cumulative and exponential growth of noise as the subgraph size increases in any publication under LDP. We address this issue by proposing an algorithm to publish the number of paths that start at every node in the graph, leading to an algorithm that publishes the Katz centrality of all nodes. This algorithm employs multiple rounds of communication and the clipping technique. Both our theoretical and experimental assessments indicate that our algorithm exhibits acceptable bias and variance, considerably less than an algorithm that bypasses clipping. Furthermore, our Katz centrality estimation is able to recall up to 90\% of the nodes with the highest Katz centrality.
\end{abstract}



\begin{document}

\maketitle

\section{Introduction}

Preserving the privacy of social network users' information is gaining in importance, especially when disclosing data or applying data mining algorithms to these networks \cite{narayanan2009anonymizing,backstrom2007wherefore,zheleva2009join}. The typical method of ensuring privacy involves the obfuscation of the original social networks or the results of data mining. Various privacy concepts have been established to ensure that these obfuscated networks or outcomes provide sufficient privacy for users. A number of these concepts, such as $k$-diversity \cite{campan2008data} and $\ell$-diversity \cite{zhou2011k}, are extensions from privacy notions designed for tabular data. 

In the realm of tabular data, differential privacy is among the most widely adopted privacy notions, as it provides a quantifiable measure of the amount of user information disclosed in a given publication, referred to as the privacy budget \cite{dwork2006differential,dwork2014algorithmic}. The broad interest in this concept comes from its relative simplicity in calculating this privacy budget, even for complex data mining operations and data publications \cite{mcsherry2009privacy}.

Numerous variations of differential privacy are presented in literature~\cite{soria2017individual,mironov2017renyi}, with local differential privacy (LDP) \cite{evfimievski2003limiting,cormode2018privacy} being one of the most prominent. In differential privacy, the default assumption is that unaltered data is aggregated at a central server, and the obfuscation is performed on this complete data. However, LDP aims to safeguard user information during its transmission to the central server. Therefore, the data obfuscation occurs locally. Because the central server does not have access to the unmodified data at any time, it is typically more challenging to apply any data mining algorithm to the data that is protected under LDP.

Edge LDP \cite{qin2017generating}, an augmentation of LDP, has been put forth specifically for the publication of social network information. Under the protection of edge LDP, it becomes hard to discern the presence of an edge or relationship within the input social network based on the published information. Multiple graph data mining algorithms \cite{hidano2022degree, sajadmanesh2021locally,ye2020lf} have been developed within the edge LDP framework, including those for subgraph counting \cite{imola2021locally,imola2022communication,imola2022differentially,hillbrand2023}.

To the best of our understanding, all existing LDP-based counting algorithms attempt to count subgraphs identifiable via the local information of a single node or a small number of nodes, such as adjacency vectors - this includes subgraphs like $k$-stars, triangles, or 4-cycles. No work, however, has been conducted on subgraphs which require consideration of adjacency vectors of multiple nodes. This is attributable to the fact that in LDP, these vectors are obfuscated independently. Despite the low probability of addition or removal of an edge from an adjacency list, the chance of obfuscation of an edge in a larger subgraph can be quite significant, which can result in a considerable error in the counted number.  

\subsection{Our Contribution}

Our contribution in this paper is as follows:
\begin{quote}
We propose an algorithm to estimate the number of paths with specific length in a social network under LDP, and apply this algorithm to provide an estimation of Katz centrality \cite{katz1953new}, a prevalent social network centrality measure. Additionally, we carry out a thorough theoretical analysis of the algorithm.
\end{quote}

Although paths with long lengths involve several nodes in the graph, we can estimate the number using multiple rounds of communications and local clipping method. Our algorithm is discussed in Section~3. 

While the utilization of the clipping method is not a new concept and has been previously employed in \cite{imola2022communication}, we are the first to offer a theoretical guarantee for multiple rounds of communications in Section~4. Here, we give upper bounds for the variance and bias of our algorithm. Both of the upper bounds are relatively small. Our analysis facilitates the proposal of the optimal parameter for the clipping. A key factor in our analysis is our assumption that only a small number of nodes possess a large degree. It is worth noting that several practical social networks, such as those adhering to the power-law degree distribution \cite{stephen2009explaining, clauset2009power}, meet our assumption.

Section~5 confirms our theoretical results through experimental validations. This section illustrates that the bias and variance in our estimation of the number of paths and Katz centrality significantly decrease compared to scenarios without clipping. Moreover, our Katz centrality estimation effectively recalls up to 90\% of the nodes with peak Katz centrality. Consequently, it provides precise recommendations of the most influential nodes in the social networks, as gauged by Katz centrality.

Calculating path counts and Katz centrality present difficulties not only to the LDP notion, but also to the general concept of differential privacy. The reason for this is that the number of paths can undergo massive shifts with the addition or removal of a single edge. This results in high sensitivity and requires the addition of substantial noise to the count in order to protect user information. We have attempted to use similar proof methods as in Section~4 to arrive at a lower upper bound on sensitivity. However, despite the improved upper bound, all the differential privacy notion algorithms we experimented with failed to surpass our Section~3 algorithm. Hence, we conclude that not only is this algorithm optimal for LDP, but it is also the best differential privacy algorithm for estimating the number of paths and Katz centrality. 

\subsection{Related Works}

The domain of graph data mining under LDP is comparatively new, whereas mining under differential privacy has been a subject of investigation for several researchers over the years \cite{gupta2010differentially,olatunji2021releasing}.  Except for special cases such as \cite{zhang2020differentially}, LDP generally only allows for the concealment of an edge or relationship \cite{imola2021locally}, while differential privacy can also be used to hide whether an individual or node is part of a social network \cite{hay2009accurate,raskhodnikova2016differentially}. In essence, there exists edge differential privacy and node differential privacy, but the concept of node differential privacy is not applicable in the context of LDP.

There are algorithms publishing centrality of graphs under differential privacy \cite{laeuchli2022analysis,task2012guide}. The most notable one is the algorithm for publishing PageRank centrality \cite{epasto2022differentially}. One might think that the publication of PageRank and Katz centrality are similar as both of them are based on the repetition of matrix multiplication. However, unlike the Katz centrality, the sensitivity of the PageRank centrality does not grow with the number of repetitions. The analysis conducted in the previous paper  therefore be applied for our task.

\section{Preliminaries}

\subsection{Notations}
In this section, we establish the notations that will be consistently used in this paper.
We represent $\mathbb{N}$ as the set of integers and $\mathbb{R}$ as the set of real numbers.
Furthermore, an input social network is denoted by $G = ([n],E)$ when $[n] = \{1,2,...,n\}$ and $E \subset [n]^{2}$. We use $\mathbb{G}_n$ to represent the collection of graphs consisting of $n$ nodes. 

For every $v \in [n]$, denote $a_{v} \in \{0, 1\}^{n}$ as the adjacency vector of node $v$. In this context, $a_{v}[u]=1$ signifies that nodes $v$ and $u$ are neighbors, otherwise it is 0.
For each vector $a \in \{0, 1\}^{n}$, let $\Gamma(a) \subset \{0,1\}^{n}$ correspond to the collection of lists that are different from $a$ by a single bit. The set $\eta(v) \subset [n]$ denotes the set of nodes adjacent to $v$ and $deg(v) < n$ denotes the degree of $v$ in the graph $G$.

Two graphs, $G = ([n],E)$ and $G' = ([n],E')$, are said to differ by a single edge if an edge $e \in [n]^{2}$ exists such that $E=E'\cup \{e\}$ or $E'=E\cup \{e\}$. The set of all graphs differing from $G$ by one edge is represented as $\Gamma(G) \subset \mathbb{G}_{n}$.

For every $\delta \in \mathbb{R}_{\geq 0}$, the Laplacian noise centered at 0 with a scale of $\delta$ is represented as $Lap(\delta)$.
For any $k > 0$ and $a \in \mathbb{R}^{k}$, the 1-norm of $a$ is denoted by $|a|$.

\subsection{Local Differential Privacy for Graph Data Mining}

Definitions of the local differential privacy (LDP) and relationship differential privacy (RDP) for graph and social network information used in this paper are as follows: 

\begin{definition}[$\epsilon$-edge LDP \cite{qin2017generating}]
For any $\epsilon \in \mathbb{R}{\geq 0}$ and $v \in [n]$, if $R_{v}$ is a randomized algorithm with the domain $\{0,1\}^{n}$, then $R_{v}$ is said to offer $\epsilon$-edge LDP if, for any pair of neighbor lists $a_{v},a'_{v}$ that only differ by a single bit and for any $S \subset range(R_{v})$, $\mathbb{P}[R_{v}(a'_{v}) \in S] \leq \exp(\epsilon) \cdot \mathbb{P}[R_{v}(a_{v}) \in S].$
\label{def:edgeLDP}
\end{definition}

\begin{definition}[$\epsilon$-edge RDP \cite{imola2021locally}]
Let $\epsilon \in \mathbb{R}_{\geq 0}$ and $G = ([n],E)$ be a graph. Let $(R_{v})_{v \in [n]}$ be a family of randomized algorithm which takes into input $a_{v}$. We say that the family of algorithm $(R_{v})_{v \in [n]}$ provides $\epsilon$-edge RDP if, for all $S \subset range(R_{1}) \times \cdots \times range(R_{n})$ and for all $G' \in \Gamma(G)$, we have that
$\mathbb{P}[(R_{1}(a'_{1}),\dots,R_{n}(a'_{n})) \in S] \leq \exp(\epsilon) 
\cdot \mathbb{P}[(R_{1}(a_{1}),\dots,R_{n}(a_{n})) \in S],$
when $a_v$ and $a'_v$ are adjacency vectors of the node $v$ in the graph $G$ and $G'$.
\label{def:edgeRDP}
\end{definition}

The upcoming theorem explores the connection between $\epsilon$-edge LDP and $\epsilon$-edge RDP. 

\begin{theorem}[Proposition~1 of \cite{imola2021locally}]
If a family of independent randomized algorithms $(R_{v})_{v \in [n]}$ provides $\epsilon$-edge LDP, then $(R_{v})_{v \in [n]}$ also provides $2 \epsilon$-edge RDP.
\end{theorem}

We also have the following property for the $\epsilon$-edge RDP.

\begin{theorem}[Composition Theorem~3.14 of \cite{dwork2014algorithmic}]
Let $\mathcal{A}_1, \dots, \mathcal{A}_S$ be a family of randomized algorithms that are $\epsilon$-edge RDP, and let $\mathcal{A}$ be an algorithm that uses only $\mathcal{A}_1, \dots, \mathcal{A}_S$ to gather information from users. Then, $\mathcal{A}$ is $(\epsilon S)$-edge RDP.
\end{theorem}

\subsection{Sensitivity and Laplacian Mechanism}

In this section, we introduce the Laplacian mechanism \cite{dwork2006calibrating}, an algorithm that generates outputs in compliance with the $\epsilon$-edge LDP and RDP as defined in the preceding subsection. We begin with the definition of sensitivity:

\begin{definition}[Sensitivity]
Let $R$ be a deterministic algorithm of which the domain is $\{0;1\}^{n}$ and the range is $\mathbb{R}^{k}$ for $k > 0$, we say that $R$ has a sensitivity of $\sigma$ if 
$\max\limits_{a \in \{0;1\}^{n},a' \in \Gamma(a)} |R(a)-R(a')| \leq \sigma$
\end{definition}

We are now ready to describe the Laplacian mechanism. 

\begin{definition}[Laplacian Mechanism \cite{dwork2006calibrating}] 
Let \newline $R_i: \{0, 1\}^{n} \rightarrow \mathbb{R}^{k}$ be a deterministic algorithm, let $\sigma_i$ be the sensitivity of $R_i$ and let $Y_i = (Y_{i1},\dots,Y_{ik})$ where the $Y_{ij}$ are drawn independently from $Lap(\sigma_i/\epsilon)$. We say that $R_i': \{0, 1\}^{n} \rightarrow \mathbb{R}^{k}$ is a publication of $R_i$ under the Laplacian mechanism if $R_i'(a_i) = R_i(a_i) + Y_i$.      
\end{definition}

The following theorem can be straightforwardly derived from Proposition 1 of \cite{dwork2006calibrating}.

\begin{theorem}  
For all $i$, let $R_i'$ be a publication of $R_i$ under the Laplacian mechanism. Then, $(R_{i})_{i \in [n]}$ provides $\epsilon$-edge LDP. \label{thm:ldprdp}
\end{theorem}

As a result of the previous theorem, we obtain that the Laplacian mechanism is $2\epsilon$-edge RDP.

\subsection{Number of Paths and Katz Centrality}

For every \( v \in [n] \) and \( k \in \mathbb{N} \), our goal is to compute the vector \( P^{(k)} \) where \( P^{(k)}[v] \) denotes the number of paths of length \( k \) originating from \( v \). We can determine \( P^{(k)}[v] \) for each \( v \in [n] \) based on the principle that, for all \( k > 0 \) and \( v \in [n] \), \( P^{(k)}(v) = \sum\limits_{u \in \eta(v)} P^{(k-1)}(u) \).

Introduced by Leo Katz in 1953, Katz centrality, also known as alpha centrality, is a widely recognized method for assessing the significance of nodes in networks. Let \( \alpha \) be a constant, referred to as the attenuation factor. For a node \( v \in [n] \), its Katz centrality is defined as $Katz[v] = \sum\limits_{k=1}^{\infty} \alpha^{k} P^{(k)}[v].$

For $i > 0$ and $v \in [n]$, let $K^{(i)}[v] = \alpha^i P^{(i)}[v]$. We have that $K^{(i - 1)}[v] = \alpha \cdot \sum\limits_{u \in \eta(v)} K^{(i)}[u]$ and $Katz[v] = \sum\limits_{k = 1}^{\infty} K^{(k)}[v]$. We then can use Algorithm~\ref{alg:Katz} to calculate $Katz[v]$ for all $v \in [n]$. As the number of step $S$ we use gets bigger, the approximation of Katz centrality gets exponentially better.

\begin{algorithm}
\SetAlgoLined
\SetKwInOut{Input}{Input}\SetKwInOut{Output}{Output}

\Input{Graph $G = ([n],E)$, attenuation factor $\alpha$, number of step $S$}
\Output{Vector $Katz$ of size $n$ where $Katz[i]$ is the Katz centrality of node $i$}
\BlankLine
$\textit{Katz} \gets \textbf{0}^{n}$, $ \textit{K}^{(0)} \gets \textbf{1}^{n}$ \;
\For{$i = 1$ \KwTo $S$}{
    \For{$v \in [n]$}{
    $K^{(i)}[v] \gets \alpha \cdot \sum\limits_{u \in \eta(v)} K^{(i - 1)}[u]$ \;
    $Katz[v] \gets Katz[v] + K^{(i)}[v]$
    }
}
\Return $Katz$
\caption{Computation of Katz centrality}
\label{alg:Katz}
\end{algorithm}

Let $A$ be the adjacency matrix of the input graph $G$. It is known that, for the series to converge, \( \alpha \) must be less than the inverse of the maximum absolute eigenvalue of \( A \). One can calculate the Katz centrality vector as $Katz = ((I - \alpha A^{T})^{-1} - I) J,$
where \( I \) is the identity matrix and \( J \) is an \( n \)-dimensional vector filled with ones. This can be computed by a single matrix inversion, but Algorithm~\ref{alg:Katz} is easier to adapt to a LDP framework.

It is important to note that $K^{(i)} = P^{(i)}$ for $i$ if we set $\alpha = 1$.

\section{Our Estimator of Katz Centrality and number of paths}

In this section, we introduce an algorithm tailored to estimate Katz centrality values in the context of LDP and RDP notions. This algorithm is detailed in Algorithm~\ref{alg:LDPKatz}. Additionally, by the section's conclusion, we give the modifications necessary to employ this algorithm for path number estimations.

The underlying principle of Algorithm~\ref{alg:LDPKatz} recognizes that while Katz centrality inherently depends on the global graph topology, the computations executed by Algorithm~\ref{alg:Katz} for each node are autonomous, relying solely on its immediate neighbors. Consequently, the algorithm can be decentralized, allowing each node to perform its computations independently. In other words, since every node \( u \in [n] \) requires \( K^{(i-1)} \) to determine \( K^{(i)}[u] \), nodes relay their results after each step to a centralized server. This server, in turn, disseminates the entire vector \( K^{(i-1)} \) to all nodes before initiating step \( i \).

To ensure differential privacy, each node \( u \) incorporates Laplacian noise into \( K^{(i)}[u] \) prior to its transmission and also before it contributes to the Katz centrality estimation (as seen in line 11). The central server remains unaware of the graph's edge details and serves solely as a communication facilitator, ensuring that our algorithm is secure under the LDP and RDP notion.

The initial version of Algorithm~\ref{alg:LDPKatz}, excluding lines 12-13, represents our preliminary design and will serve as a comparison standard in our Section 5 experiments. However, this iteration presents an inherent flaw. The magnitude of the Laplacian noise must align with the sensitivity, denoted as \(\max\limits_{v \in [n]} |\widetilde{K}^{(i-1)}[v]|\). This becomes problematic as this magnitude can escalate considerably, potentially compromising the accuracy of our estimator from both theoretical and practical perspectives.

Motivated by \cite{epasto2022differentially}, we incorporated a clipping strategy (as presented in lines 12-13). By constraining \(K^{(i)}\) during the \(i\)th step, we aim to diminish the sensitivity when deriving \(K^{(i+1)}\) in the subsequent step. This adaptation holds potential to enhance the estimator's efficacy by minimizing noise and, consequently, variance, though it might introduce a certain bias. It is important to clarify that this clipping is executed subsequent to the incorporation of \(K^{(i)}\) into \(\widetilde{Katz}\) (as detailed in line 11). The primary intent behind the clipping is not the preservation of differential privacy during the \(i\)-th phase but rather the attenuation of sensitivity for the \(i+1\) step.

\begin{algorithm}
\SetAlgoLined
\SetKwInOut{Input}{Input}\SetKwInOut{Output}{Output}

\Input{Graph $G = ([n],E)$, attenuation factor $\alpha$, clipping factor $X$, privacy budget $\epsilon$, number of step $S$}
\Output{Vector $\widetilde{Katz}$ of size $n$ where $\widetilde{Katz}[v]$ is the estimated Katz centrality of node $v \in [n]$ under $\epsilon$-edge LDP}
\BlankLine
\For{$v \in [n]$}{
\textbf{[User $v$]} $\widetilde{Katz}[v] \gets 0$\;
\textbf{[User $v$]} $\widetilde{K}^{(0)}[v] \gets 1$ \;
}
\For{$i = 1$ \KwTo $S$}{
    \textbf{[Server]} $Noise \gets \frac{2 \alpha S}{\epsilon} \cdot \max\limits_v 
    |\widetilde{K}^{(i - 1)} [v]|$ \;
    \textbf{[Server]} Distribute the value of $Noise$ and $\widetilde{K}^{(i - 1)}$ to all users \;
    \For{$v \in [n]$}{
    \textbf{[User $v$]} $\widetilde{K}^{(i)}[v] \gets \alpha \cdot \sum\limits_{u \in \eta(v)} \widetilde{K}^{(i - 1)}[u]$\;
    \textbf{[User $v$]} $\widetilde{K}^{(i)}[v] \gets \widetilde{K}^{(i)}[v] + Lap(Noise)$ \;
    \textbf{[User $v$]} $\widetilde{Katz}[v] \gets \widetilde{Katz}[v] + \widetilde{K}^{(i)}[v]$\;
    \textbf{[User $v$]} $\widetilde{K}^{(i)}[v] \gets \min\{\widetilde{K}^{(i)}[v], (\alpha X)^i\}$ \;
    \textbf{[User $v$]} $\widetilde{K}^{(i)}[v] \gets \max\{\widetilde{K}^{(i)}[v], -(\alpha X)^i\}$ \;
    \textbf{[User $v$]} Communicate $\widetilde{K}^{(i)}[v]$ to the central server.
    }
}
\Return $\widetilde{Katz}$
\caption{Algorithm to approximate Katz centrality under \( \epsilon \)-edge LDP}
\label{alg:LDPKatz}
\end{algorithm}

From the following lemma, we show that Algorithm~\ref{alg:LDPKatz} is $\epsilon$-edge RDP. We begin by discussing the privacy of the communication at Line 14 of the algorithm.

\begin{lemma}
The communication of $\widetilde{K}^{(i)}[v]$ at Line 14 of Algorithm~3 is ($\epsilon/(2S)$)-edge LDP. \label{lem:pri}
\end{lemma}
\begin{proof}
Consider $a_v$ and $a_v'$ as adjacency vectors of node $v$, differing by a single bit. Let $\widetilde{K}_{\text{max}} := \max\limits_v |\widetilde{K}^{(i-1)}[v]|$, and suppose $\widetilde{K}[v]$ and $\widetilde{K}'[v]$ are the computation results acquired from Line 9 of the algorithm when the adjacency vector is $a_v$ and $a_v'$. We find that $|\widetilde{K}[v] - \widetilde{K}'[v]| \leq \alpha \cdot \widetilde{K}_{\text{max}}$. Therefore, the sensitivity of transmitting $\widetilde{K}^{(i)}[v]$ is $\sigma = \alpha \cdot \widetilde{K}_{\text{max}}$. Given that we set the Laplacian noise parameter to $\frac{2 \sigma S}{\epsilon}$, the Laplacian mechanism at Line 10 exhibits $\epsilon/(2S)$-edge LDP. Any post-processing not related on the edge set will not alter the privacy outcome. Hence, the communication at Line 13 ensures ($\epsilon/(2S)$)-edge LDP, despite the post-processing at Lines 12-13 of the algorithm.
\end{proof}

We are now ready to show the privacy of Algorithm~2 in the following theorem. 

\begin{theorem}
Algorithm 2 is an $\epsilon$-edge RDP. \label{thm:pri}
\end{theorem}

\begin{proof}
Using Lemma \ref{lem:pri} in conjunction with Theorem \ref{thm:ldprdp}, it is established that the $\widetilde{K}^{(i)}[v]$ communication at the algorithm's Line~14 adheres to $(\epsilon/S)$-edge RDP. Since there are $S$ round of publications, Theorem~2.2 confirms that Algorithm~3 is $\epsilon$-edge RDP.
\end{proof}

If we set $\alpha = 1$, Algorithm~\ref{alg:LDPKatz} gives us an estimator of the number of paths where $\widetilde{K}^{(i)}$ is an estimator of $P^{(i)}$.

\section{Theoretical analysis of our algorithms}

In the following discussion, we conduct an evaluation of the accuracy of the algorithm proposed in the previous section. We observed that in the majority of social networks, a handful of nodes exhibit a considerably larger degree than the rest, as affirmed several works including \cite{stephen2009explaining, clauset2009power}. This observation motivates our assumption in the following analysis. Here, we assume that the maximum degree of the input graph $G = ([n],E)$ is $D$, and there is at most $N \ll n$ nodes exhibit a degree greater than $d \ll D$.

Let us revisit the clipping factor $X$ as outlined in the preceding section. For this section, we select parameters $d$ and $X$ such that they satisfy the conditions $NX + Dd \leq X^2$ and $X \leq D$. As previously stated, we operate under the assumption that $N$ and $d$ is small, thus intuitively setting $X$ to $O\left(\sqrt{D}\right)$. 

It is noteworthy that it is always possible to identify parameters $d$ and $X$ that fulfil the condition, specifically when $X = d = D$. This results in $N = 0$, and consequently, both of the conditions are satisfied. However, by assigning $X = D$, the computation at Line 11 of Algorithm~2 becomes nearly insignificant as $\widetilde{K}^{(i)}[v]$ is typically less than $(\alpha D)^i$. For the computation to be utilized effectively, we generally aim to set the parameter of $X$ to the lowest possible value. The most ideal situation is when we can assign $X$ a value approximately equal to $\sqrt{D}$.

\subsection{Bias of Algorithm~3}

In this section, we give an upper bound of the bias of Algorithm~3 as an estimator of Katz centrality. 

\subsubsection{Main Theorem} Let $\phi$ be the golden ratio. The main results of this section are as follows:

\begin{theorem}
\label{thm:biaspath} For Algorithm~3, when considering an attenuation factor of \(\alpha=1\), a number of steps \(S\) such that \(S \geq i\), and satisfying the condition \(X^{2}/D + X \leq X^{2}\), the bias of the estimator for the number of paths of length \(i\) is given by:
\[
\max_{v \in [n]} \mathbb{E}[P^{(i)}[v]-\widetilde{K}^{(i)}[v]] \leq 2 (\phi X)^{i-1} D \frac{S}{\epsilon}.
\]
\end{theorem}

Given that the number of paths of length \(i\) originating from node \(v\) can scale as \(D^i\). When \(X \ll D\), it is evident that the upper bound presented in Theorem \ref{thm:biaspath} is significantly less than the trivial upper bound. Thus, while the clipping in lines 12-13 does introduce a certain level of bias, this bias is not large compared to the genuine number of paths.

\begin{theorem} \label{thm:bias}
 The bias of the Katz centrality estimator in Algorithm~3 with attenuation factor $\alpha < 1/ \phi X$, number of step $S$, and privacy budget $\epsilon$ such that $S/\epsilon \geq 1$  and $X^{2}/D + X \leq X^{2}$ is : 
\[\max_{v \in V} \mathbb{E}[Katz(G)[v] - \widetilde{Katz}(G)[v]] \leq \frac{\alpha S}{\epsilon}\left(1 + \frac{2 \alpha \phi D X}{1- \alpha \phi X}\right).\]
\end{theorem}

For values of \(\alpha\) less than \(1/2\phi X\), the expression \(\alpha\left(1 + \frac{2 \alpha \phi D X}{1- \alpha \phi X}\right)\) approaches a minimal constant. Consequently, the bias of the Katz centrality estimator becomes linearly proportional to \(S/\epsilon\). This suggests that, for those $\alpha$, our bias does not escalate rapidly with additional steps in Algorithm~\ref{alg:LDPKatz} and with a better level of differential privacy. 

\subsubsection{Proofs} 

Firstly, it is evident that \(0 \leq \mathbb{E}[\widetilde{K}^{(i)}[v]] \leq K^{(i)}[v]\) for all \(i > 0\) and \(v \in [n]\). This implies that both \(\widetilde{K}^{(i)}\) and \(\widetilde{Katz}\) possess a negative bias relative to the actual values. Given that the clipping at line 13 elevates \(\mathbb{E}[\widetilde{K}^{i}[v]]\), it reduces the bias (in terms of magnitude). As our objective is to present an upper boundary for this bias, we can disregard the effect of line 13.

The bias introduced by the Laplacian mechanism at Line 10 of Algorithm~\ref{alg:LDPKatz} is not easy to analyze. To facilitate analysis, we make a substitution throughout all analyses in this section. Specifically, we replace the Laplacian distribution with an alternative one that simplify our analysis and always yields a greater bias. This substitution allows us to establish an upper bound for the bias from the algorithm.

Recall that we draw the noise from the Laplacian distribution $Lap\left[\frac{2 \alpha S}{\epsilon} \max\limits_{v} \textit{K}^{(i-1)}\right]$ in Algorithm~2. Let $\mathcal{L}_{i,v}$ be the noise we have drawn. The noise is clipped to $\min\{\mathcal{L}_{i,v}, (\alpha X)^i - \widetilde{K}^{(i)}[v] \}$ at Line 12 of the algorithm. It is straightforward to see that the clipped noise does not introduce more bias than $\min\{\mathcal{L}_{i,v}, 0\}$. Hence, to facilitate the proof in this section, we assume that the alternative noise is obtained by the Laplacian distribution clipped by $0$.
In other words, our noise $\zeta_{i,v}$ is drawn from the distribution $-Exp[\frac{\epsilon}{2S\alpha} /(\alpha^{i-1} X^{i-1})]$ with probability 0.5 and is $0$ with probability 0.5. We know that $\zeta_{i,v} \leq 0$.
With the alternative noise, we obtain the following lemma:

\begin{lemma} For all $i \geq 2$ and $v \in V$, $\max\limits_{v \in V}\widetilde{K}^{(i)}[v] < (\alpha X)^{i}.$ \label{lem:largestK}
\end{lemma}
\begin{proof}
To prove this lemma, we proceed by induction on the number of step $i$. 

Define $M_{i} := \max\limits_{v \in V}\widetilde{K}^{(i)}[v]$ and 
$m_{i} = \max\limits_{v \in V : deg(v) \leq d}\widetilde{K}^{(i)}[v]$.
After the calculation at Line 12 of Algorithm~2, we have $M_{1} \leq \alpha X$ and $m_{1} \leq \alpha d$. Therefore, for all $v \in V$,
\[\begin{split}
\widetilde{K}^{(2)}[v] & = \sum_{u \in \eta(v)} \alpha \widetilde{K}^{(1)}[u] + \zeta_{i,v} \\
& \leq \sum_{u \in \eta(v) \mid deg(u) > d} \alpha \widetilde{K}^{(1)}[u] + 
\sum_{u \in \eta(v) \mid deg(u) \leq d} \alpha \widetilde{K}^{(1)}[u] \\
& \leq \alpha N M_{1} + \alpha D m_{1} \leq \alpha^{2} (N X + D d) \leq (\alpha X)^{2}
\end{split}\]
This proves that $M_{2} = \max\limits_{v \in V}\widetilde{K}^{2}[v] \leq (\alpha X)^{2}$. 

Then, for all $i > 1$ and $v \in \{u \in [n]: deg(u) \leq d\}$, we have that 
$\widetilde{K}^{(i)}[v] = \sum_{u \in \eta(v)} \alpha \widetilde{K}^{(i-1)}[u] + \zeta_{i,v} \leq \alpha d M_{i-1}$. 
This means that $m_{i} \leq \alpha d M_{i-1}$. 

To show the induction step,  we assume that $\widetilde{K}^{(j)}[v] < \alpha^j X^j$ for all $j < i$ and $v \in [n]$. By the assumption, for all $i > 2$ and $v \in [v]$: 
\[\begin{split}
\widetilde{K}^{(i)}[v] & = \sum_{u \in \eta(v)} \alpha \widetilde{K}^{(i-1)}[u] + \zeta_{i,v} \\
& \leq \sum_{u \in \eta(v) \mid deg(u) > d} \alpha \widetilde{K}^{(i-1)}[u] + 
\sum_{u \in \eta(v) \mid deg(u) \leq d} \alpha \widetilde{K}^{(i-1)}[u] \\
& \leq \alpha N M_{i-1} + \alpha D m_{i-1}  \leq \alpha N M_{i-1} + \alpha^{2} d D M_{i-2} \\
& \leq \alpha N (\alpha X)^{i-1} + \alpha^{2} d D (\alpha X)^{i - 2} \leq \alpha^iX^{i - 2} (NX + d D ) \leq (\alpha X)^{i}
\end{split}\]
\end{proof}

The previous lemma implies that, by the alternative noise used in this section, the calculation at Line 12 changes the results only at the first step. We are now ready to prove our main theorem.

\begin{proof}[Proof of Theorem \ref{thm:bias}]
The expected value of the alternative noise is $\mathbb{E}[\zeta_{i,v}] = - \alpha^{i} X^{i-1} S/\epsilon$. We obtain that, for $i > 1$, $\mathbb{E}[{K}^{(i)}[v] - \widetilde{K}^{(i)}[v]] = 
\sum_{u \in \eta(v)} \alpha \mathbb{E}[{K}^{(i-1)}[u] - \widetilde{K}^{(i-1)}[u]] + 
\frac{\alpha^{i} X^{i-1} S}{\epsilon}$.
Let $b_{i} = \max\limits_{v \in V \mid deg(v) \leq d} \mathbb{E}[{K}^{(i)}[v] - \widetilde{K}^{(i)}[v]]$
and  $B_{i} = \max\limits_{v \in V} \mathbb{E}[{K}^{(i)}[v] - \widetilde{K}^{(i)}[v]]$. From Lemma \ref{lem:largestK}, we can obtain that, for all $i > 1$, 
$B_{i} \leq \alpha (N B_{i-1} + D b_{i-1}) + \alpha^{i} X^{i-1} S/\epsilon$ and
$b_{i} \leq \alpha d B_{i-1} + \alpha^{i} X^{i-1} S/\epsilon$. Therefore, for all $i > 2$, 
$B_{i} \leq \alpha N B_{i-1} + \alpha^{2} d D B_{i-2} + 
\alpha^{i} (X^{i-1} + D X^{i-2}) \frac{S}{\epsilon}.$
We can now prove by induction that $B_{i} \leq 2 \alpha^{i} (\phi X)^{i-1} D S/\epsilon$. By the assumption that $S/\epsilon \geq 1$, we have:  $B_{1} \leq \alpha (D-X) + \alpha S/\epsilon
\leq 2 \alpha D S/\epsilon$.
Recall the assumption that $NX + Dd \leq X^2$ and $X \leq D$. The assumptions imply that $N \leq X$ and $d \leq X$. It follows that $b_1 \leq \alpha S / \epsilon$. Hence, by $\alpha < 1/\phi X$:
\begin{dmath*}B_{2} \leq \alpha (N B_{1} + (D-N) b_{1}) + \alpha^{2} X \frac{S}{\epsilon} \hiderel{\leq} \alpha^2 N(D - X) + \alpha^2 D \frac{S}{\epsilon} + \alpha^2 X \frac{S}{\epsilon}  \leq \alpha^{2} N (D-X) + \alpha^{2} (D+X) S/\epsilon. \end{dmath*} 
Recall that $N \leq X$ and $1 \leq S/\epsilon$. We obtain that $N(D-X) \leq ND \leq XDS/\epsilon$. Also, recall the assumption that $X^2/D + X \leq X^2$. We obtain that $X + D \leq XD$. Hence, 
$B_2 \leq \alpha^{2} X D S/\epsilon + \alpha^{2} X D S/\epsilon \leq 2 \alpha^{2} \phi X D S/\epsilon$.

We will now consider the case when $i \geq 3$. Assume by induction that for all $k < i$, $B_{k} \leq 2\alpha^{k} (\phi X)^{k-1} D S/\epsilon$, then 
\[
\begin{split}
B_{i} & \leq \alpha N B_{i-1} + \alpha^{2} d D B_{i-2} +  \alpha^{i} (X^{i-1} + D X^{i-2}) S/\epsilon \\
& \leq \frac{2 \alpha^{i} S D (\phi X)^{i-3}}{\epsilon} \left[ N \phi X + d D + \frac{(X^{2}+D X)}{D \phi^{i-2}}\right] \\
& \leq \frac{2 \alpha^{i} S D (\phi X)^{i-3}}{\epsilon}( N (\phi-1) X + 
(N X + d D) + (X^{2}/D + X)) \\
& \leq \frac{2 \alpha^{i} S D (\phi X)^{i-3}}{\epsilon}( (\phi-1) X^2 + 
X^2 + X^2) \\
& \leq \frac{2 \alpha^{i} S D \phi^{i-3} X^{i-1}}{\epsilon}( \phi + 1) = 
\frac{2 \alpha^{i} S D (\phi X)^{i-1}}{\epsilon}
\end{split}
\]
This concludes the induction. 

The discrepancy in the Katz centrality estimated by our algorithm is from three components:
\begin{enumerate}
\item The bias derived from the initial step, which does not exceed $\alpha S/\epsilon$;
\item The bias from the second up to the $S$-th step, which is not larger than the sum $\sum\limits_{i = 2}^S B_i$; and
\item The bias resulting from the limitation that our computation does not extend past the $S$-th calculation step.
\end{enumerate}
Therefore, 
\[
\begin{split}
& \max_{v \in V} \mathbb{E}[Katz(G)[v] - \widetilde{Katz}(G)[v]] \leq \frac{\alpha S}{\epsilon} + 
\sum_{i=2}^{S} B_{i}  + \sum_{i= S + 1}^{\infty} K^{(i)}[v] \\
& \leq \frac{\alpha S}{\epsilon} +
\sum_{i=2}^{S} \frac{2 \alpha D S (\alpha \phi X)^{i-1}}{\epsilon}
+ \sum_{i=S + 1}^{\infty} (\alpha X)^{i}
\\
& \leq \frac{\alpha S}{\epsilon} + \frac{2 \alpha^{2} D S \phi X}{\epsilon} \sum_{i=0}^{S-2} (\alpha \phi X)^{i} + \frac{(\alpha X)^{S+1}}{1-\alpha X} \\
& \leq \frac{\alpha S}{\epsilon} + 
\frac{2 \alpha^{2} D S \phi X(1 - (\alpha \phi X)^{S-1})}{\epsilon(1- \alpha \phi X)} + 
\frac{(\alpha X)^{S+1}}{1-\alpha X} \\
& \leq \frac{\alpha S}{\epsilon}(1 + \frac{2 \alpha \phi D X}{1- \alpha \phi X}) -
\frac{2 \alpha D S (\alpha \phi X)^{S}}{\epsilon (1 - \alpha \phi X)} +
\frac{(\alpha X)^{S+1}}{1-\alpha X}.
\end{split}
\]
By $2S/\epsilon > 1$, $X \leq D$, and $\phi > 1$, we obtain the theorem statement from the previous derivation because 
$\frac{2 \alpha D S (\alpha \phi X)^{S}}{\epsilon (1 - \alpha \phi X)} \geq \frac{(\alpha X)^{S+1}}{1-\alpha X}.$
\end{proof}

The proof of Theorem \ref{thm:biaspath} is a result from derivations in the previous proof. The theorem's assertion can be derived by assigning a value of \( \alpha \) in the previous proof's upper bound of \( B_i \) to be \( 1 \).

\subsection{Variance of Algorithm~\ref{alg:LDPKatz}}

\subsubsection{Theorems}
Let $L = \max(N D,X^{2})$. The following theorems provide the upper bound of the variance for the number of paths and Katz centrality that Algorithm~\ref{alg:LDPKatz} publishes.

\begin{theorem} \label{thm:pathvariance}
For the estimator of the number of paths of length \( i \) obtained from Algorithm~\ref{alg:LDPKatz}, given an attenuation factor \( \alpha=1 \), number of steps \( S\geq i \), and satisfying the condition \( X^{2}/D+X \leq X^{2} \), the variance is bounded by:
\[\max_{v \in [n]} \mathrm{Var}[\widetilde{K}^{(i)}[v]] \leq \frac{32 S^{2} (D^{2} + X^{2}) (4 L)^{i-2}}{\epsilon^{2}}.\]
\end{theorem}

The theorem suggests that the standard deviation of our publication scales as \( (2X)^{i - 1} \). Given that \( X \ll D \) and the typical path can scale as \( D^i \), the upper bound of the standard deviation is not large compared to the actual path count.

\begin{theorem} \label{thm:katzvariance}
The variance of the Katz estimator published by Algorithm~\ref{alg:LDPKatz} using attenuation factor $\alpha \leq 1/(2\sqrt{L})$, number of step $S$ and privacy budget $\epsilon$ is:
\[\max_{v \in V} \mathrm{Var}[\widetilde{Katz}(G)[v]] \leq \frac{8 S^{2} \alpha^{2}(D^{2} + X^{2})}{ L \epsilon^{2} (1 - 2 \alpha \sqrt{L})^{2}} \]
\end{theorem}

For values of \( \alpha \) such that \( \alpha \leq \frac{1}{4\sqrt{L}} \), the term \( \frac{8 \alpha^2 (D^2 + X^2)}{L(1 - 2\alpha\sqrt{L})^2} \) tends toward a negligible constant. As a result, the standard deviation of the Katz estimator produced by our algorithm aligns with the order of \( \frac{S}{\epsilon} \). This observation  indicates that, for these values of \( \alpha \), the variance remains relatively stable even as the step count \( S \) and privacy parameter \( \epsilon \) vary.

\subsubsection{Proofs}

\begin{proof}[Proof of Theorem \ref{thm:katzvariance}]
First, let us examine the variance of the number of paths, $K^{(i)}[v]$. It is clear that the computation at Lines~12-13 in Algorithm~\ref{alg:LDPKatz} can only decrease the variance, so we can disregard this step when establishing the upper bound for the variance. Consequently, the upper bound for $Var[K^{(i)}[v]]$ is made up of two components:\\
(1) The variance of the Laplacian noise added at Line 10: By Lemma \ref{lem:largestK}, we have that $Noise \leq \frac{2\alpha S}{\epsilon} \cdot \max\limits_{v \in V}  \widetilde{K}^{(i-1)}[v] < \frac{2 \alpha^{i}X^{i-1} S}{\epsilon}.$ Hence, the variance of the Laplacian noise at Line 10 is not larger than $2 \cdot Noise^2 = 8 \alpha^{2 i} X^{2 i -2} S^{2}/\epsilon^{2}$. \\
(2) The collective sum of covariances between the variable $\alpha \cdot \widetilde{K}^{(i-1)}[u]$ and $\alpha \cdot \widetilde{K}^{(i-1)}[w]$ for every $u,w$ within $\eta(v)$: Let  $\mathcal{V}_i = \max\limits_{v\in V}\mathrm{Var}[\widetilde{K}^{(i)}[v]]$ and 
$\nu_i = \max\limits_{v\in V : deg(v) \leq d}\mathrm{Var}[\widetilde{K}^{(i)}[v]]$. By the Cauchy-Schwartz inequality \cite{cauchy1821cours}, we obtain that
\begin{equation*}
    \mathrm{Cov}[\widetilde{K}^{(i)}[u], \widetilde{K}^{(i)}[w]] \leq 
    \begin{cases} 
      \nu_i & \text{if } deg(u), deg(w) \leq d, \\
      \sqrt{\nu_i \cdot \mathcal{V}_i} & \text{if } \min\{deg(u),deg(w)\} \leq d, \\ 
      \mathcal{V}_i & \text{Otherwise.}
   \end{cases}
\end{equation*}
Given that the maximum number of nodes in $\eta(v)$ is $D$, and among these $D$ nodes, at most $N$ nodes have a degree exceeding $d$:
\begin{dmath*}
    \sum_{u,v \in \eta(v)} \mathrm{Cov}[\alpha \cdot \widetilde{K}^{(i)}[u], \alpha \cdot \widetilde{K}^{(i)}[w]] 
    \leq \alpha^{2} (N^{2} \mathcal{V}_{i-1} + 
2 N D \sqrt{\nu_{i-1}\mathcal{V}_{i-1}} + D^{2} \nu_{i-1}).
\end{dmath*}
There is no need to account for the covariance between $\widetilde{K}^{(i)}[u]$ and the Laplacian noise, since the noise is generated independently of the value of $\widetilde{K}^{(i)}[u]$.

 Hence, for all $i > 1$,
\begin{align*}
\mathcal{V}_i & \leq 8 \alpha^{2 i} X^{2 i -2} S^{2}/\epsilon^{2} + \alpha^{2} (N^{2} \mathcal{V}_{i-1} + 
2 N D \sqrt{\nu_{i-1}\mathcal{V}_{i-1}} + D^{2} \nu_{i-1}), \\
    \nu_i & \leq 8 \alpha^{2 i} X^{2 i -2} S^{2}/\epsilon^{2} + \alpha^{2} d^{2} \mathcal{V}_{i-1}.
\end{align*}

By combining the above inequalities and by $\nu_i \leq \mathcal{V}_i$, we have that, for all $i >2$, 
\[
\mathcal{V}_{i} \leq \alpha^{2}(N^{2} + 2 N D) \mathcal{V}_{i-1} + \alpha^{4} d^{2} D^{2} \mathcal{V}_{i-2} 
+ \frac{8 \alpha^{2 i} X^{2 i - 4} S^{2}}{\epsilon^{2}}(D^{2} + X^{2}).
\]
By defining $L = \max\{X^{2},ND\}$, we will now prove by induction that, for all $i \geq 1$,
$\mathcal{V}_{i} \leq \frac{32 S^{2} \alpha^{2 i}}{\epsilon^{2}} (D^{2} + X^{2}) (4 L)^{i-2}$.

First, because the variance at the first step $\mathcal{V}_1$ comes only from the Laplacian noise, we have that 
$\mathcal{V}_{1} = \frac{8 \alpha^{2} S^{2}}{\epsilon^{2}} \leq \frac{32 \alpha^{2} S^{2} (D^{2}+ X^{2})}{\epsilon^2 (4L)}.$
By that, the covariance sum at the second step is no more than $\alpha^2 D^2 \mathcal{V}_1 = 8\alpha^4S^2D^2/\epsilon^2$. Because the variance of the Laplacian noise is $8\alpha^4X^2S^2 / \epsilon^2$, we have that 
$\mathcal{V}_{2} = \frac{8 \alpha^{4} S^{2} (D^{2} + X^{2})}{\epsilon^{2}} < \frac{32 \alpha^{4} S^{2} (D^{2} + X^{2})}{\epsilon^{2}}.$ 
For $i > 2$, we assume that, for all $k < i$,
$\mathcal{V}_{k} \leq \frac{32 S^{2} \alpha^{2 k}}{\epsilon^2} (D^{2} + X^{2}) (4 L)^{k-2}$, then, by $N^2 \leq X^2 \leq L$, $dD \leq X^2 \leq L$, and $ND \leq L$: 
\[
\begin{split}
\mathcal{V}_{i} & \leq \frac{32 S^{2} \alpha^{2 i} (4L)^{i-4} (D^{2} + X^{2})}{\epsilon^{2}}(4 L (N^{2} + 2 N D) + d^{2} D^{2} +X^{4}/4^{i-3}) \\
& \leq \frac{32 S^{2} \alpha^{2 i} (4L)^{i-4} (D^{2} + X^{2})}{\epsilon^{2}}(12 L^{2} + L^2 + L^{2}) \\
& \leq \frac{32 S^{2} \alpha^{2 i}}{\epsilon^{2}} (D^{2} + X^{2}) (4 L)^{i-2}.
\end{split}
\]
Finally, considering that \( \widetilde{Katz}(G)[v] = \sum_{i=1}^{S} K^{i}[v] \), and leveraging the Cauchy-Schwartz inequality, combined with the understanding that \( \sum_{i=0}^{\infty} \sum_{j=0}^{\infty} x^{i+j} = \frac{1}{(1-x)^2} \) for all \( x \) in \( \mathbb{R} \), we deduce the following:
\[
\begin{split}
\mathrm{Var}[\widetilde{Katz}[v]] & = \sum_{i=1}^{S} \sum_{j=1}^{S} \mathrm{Cov}[K^{(i)}[v],K^{(j)}[v]]\\
& \leq \sum_{i=1}^{S} \sum_{j=1}^{S} \sqrt{\mathrm{Var}[K^{(i)}[v]]\mathrm{Var}[K^{(i)}[v]]} \\
& \leq \frac{32 S^{2} \alpha^{2}}{4 L \epsilon^{2}} (D^{2} + X^{2}) 
\sum_{i=0}^{S-1} \sum_{j=0}^{S-1} (2 \alpha \sqrt{L})^{i} (2 \alpha \sqrt{L})^{j} \\
& \leq  \frac{32 S^{2} \alpha^{2}}{4 L \epsilon^{2}} (D^{2} + X^{2}) 
\sum_{i=0}^{\infty} \sum_{j=0}^{\infty} (2 \alpha \sqrt{L})^{i} (2 \alpha \sqrt{L})^{j} \\
& \leq \frac{8 S^{2} \alpha^{2}(D^{2} + X^{2})}{L \epsilon^{2} (1 - 2 \alpha \sqrt{L})^{2}} 
\end{split}
\]
\end{proof}

We can show Theorem \ref{thm:pathvariance} by the upper bound of $\mathcal{V}_i$ provided in the previous proof. 

\subsection{Variance of the Algorithm without Clipping}

We will now proceed to evaluate the algorithm devoid of clipping, essentially examining Algorithm 3 while omitting lines 12 and 13. Given that the bias of this algorithm converges to 0 as the number of steps \( S \) approaches \(\infty\), our analysis will predominantly concentrate on its variance. To streamline the evaluation, we will explore its variance over the graph \( G_0 = ([n],\emptyset) \) — a representation with \( n \) nodes and devoid of edges. Our theorem indicates that, even with this simple graph structure, the variance amplifies at such a rate that the utility of the publication becomes questionable.

Let \(N_{i}\) represent the scale of the noise at the \(i^{th}\) step, as seen in line 6 of Algorithm~\ref{alg:LDPKatz}. The underlying principle here is that when drawing \(n\) Laplacian noises of scale \(N_{i}\), it is highly probable that one of them will be considerably large, causing \(N_{i+1}\) to also be large. If we consider \((L_{i}(v))_{v \in [n]}\) as the \(n\) Laplacian noises drawn at the \(i^{th}\) step, each with a scale of \(N_{i}\), and given that \(G_0\) lacks any edges, we can deduce that \(K^{(i)}[v] = L_{i}(v)\). This leads to the expression 
$N_{i+1} = \frac{2 \alpha S}{\epsilon} \max\limits_{v \in [n]} |L_{i}(v)|$.
We will employ the subsequent lemma for further analysis:

\begin{lemma} [\cite{maxexpo}]
Let $n>0, \delta>0 $ and $(L_{v})_{v \in [n]}$ be n independent Laplacian noise with scale $\delta$. Then $\mathbb{E}[ \max\limits_{v \in [n]} |L_{v}|] = \delta H_{n}$ where $H_{n}$ is the Harmonic series with n terms.
\end{lemma}

Given the algorithm without clipping applied to graph \(G_0\), we can now determine the expected noise at step \(i\).

\begin{theorem}
Considering the algorithm without clipping, and given parameters \(G_0\), \(\alpha\), \(X\), \(S\), and \(\epsilon\). Define \(H_{n}\) as the Harmonic series with \(n\) terms, where \(n\) represents the total number of nodes. The anticipated noise for step $i$ is expressed as $(2 \alpha S/\epsilon)^{S} H_{n}^{S-1}$. \label{thm:noclip}
\end{theorem}

\begin{proof}
We proceed with the proof of the theorem using induction. For the base case, consider step 1. Given that \(\widetilde{K}^{(0)}[v] = 1\) for every \(v \in [n]\), it follows that \(\mathbb{E}[N_{1}] = 2 \alpha S/\epsilon\).

Now, let us assume for some arbitrary step \(i>0\) that \(\mathbb{E}[N_{i}] = (2 \alpha S/\epsilon)^{i} H_{n}^{i-1}\). Given that \(N_{i+1} = \frac{2 \alpha S}{\epsilon} \max\limits_{v \in [n]} |L_{i}(v)|\) where each \(L_{i}(v)\) represents an independent Laplacian noise with scale \(N_{i}\), and by employing Lemma 4.4, we can express \(\mathbb{E}[N_{i+1}|N_{i}]\) as \(\frac{2 \alpha S}{\epsilon} H_{n} N_{i}\). Consequently, we have 
$\mathbb{E}[N_{i+1}] = \mathbb{E}[\mathbb{E}[N_{i+1}|N_{i}]] = \frac{2 \alpha S}{\epsilon} H_{n} \mathbb{E}[N_{i}] = (2 \alpha S/\epsilon)^{i+1} H_{n}^{i}$.
This establishes the induction hypothesis, completing the proof.
\end{proof}

In Theorems~\ref{thm:pathvariance} and \ref{thm:katzvariance}, the variance increases linearly with $S/\epsilon$. In contrast, Theorem~\ref{thm:noclip} shows an exponential growth in noise. This suggests that when the condition $(2 \alpha S)/\epsilon > 1$ holds, the unclipped variant will have considerably greater variance. Even though clipping introduces a bias, its presence greatly diminishes the algorithm's variance. As a result, our algorithm is anticipated to perform notably better than its unclipped counterpart.

\section{Experiments}

In this section, we enhance our theoretical insights with empirical evaluations of our algorithm. Specifically, we examine its efficacy in estimating the Katz centrality and the number of paths.

\noindent \textbf{Dataset : } We conducted our experiments using two graphs sourced from the Stanford Network Analysis Platform (SNAP). The first graph represents the social circles from Facebook, as described in \cite{leskovec2012learning}. This undirected graph encompasses 4,039 nodes and 88,234 edges. Its average degree stands at 43.69, with the highest degree reaching 1045, and a maximum eigenvalue of $E_{F} = 162.37$. The second graph illustrates voting patterns on Wikipedia, based on references \cite{leskovec2010signed,leskovec2010predicting}. This directed graph has 7,115 nodes and 103,663 edges. Its average degree is 14.57, with a peak degree of 1167, and its maximum eigenvalue is $E_{W} = 45.14$.

\noindent \textbf{Baseline : } To our understanding, this work pioneers the LDP technique for estimating number of paths and Katz centrality. For benchmarking purposes, both for Katz centrality and the number of paths, we employ our algorithm, excluding the clipping mechanism (specifically, Algorithm~\ref{alg:LDPKatz} without lines 12-13). This comparative analysis aims to ascertain the tangible impact of the clipping concept on the efficiency of our method.

\noindent \textbf{Privacy Budget $\epsilon$ Setting:} Throughout our experimental evaluations, we adopted a privacy budget set at $\epsilon = 1$, a commonly accepted benchmark \cite{imola2021locally}. Based on our theoretical findings, we anticipate analogous outcomes for different values of $\epsilon$.
 
\subsection{Results for Katz Centrality}

\noindent \textbf{Attenuation Factor:} As discussed in section~2.4, the reciprocal of the graph's maximum eigenvalue acts as an upper limit for the attenuation factor. We opted for attenuation factors near this threshold, setting $\alpha_{F} = 0.85/E_{F}$ and $\alpha_{W} = 0.85/E_{W}$. Such values render the Katz centrality estimation more challenging. Specifically, for small values of $\alpha$, the relation $Katz[v] \approx \alpha \times deg(v)$ holds true for all nodes $v \in [n]$.

Firstly, we have tested the performance of our algorithms by computing by computing their loss and variance as estimators of the Katz centrality. The results are shown on \textbf{figure 1}. For \textbf{figure 1.a} and \textbf{1.c}, we used a clipping factor of $X = E_{F}$ and $X = E_{W}$ respectively and for \textbf{figure 1.b} and \textbf{1.d} we used a number of steps of $S=5$.

\begin{figure}
  \centering \vspace{-0.7cm}
  \subfloat[Varying the number of steps $S$ on the Facebook graph]{\includegraphics[width=4cm]{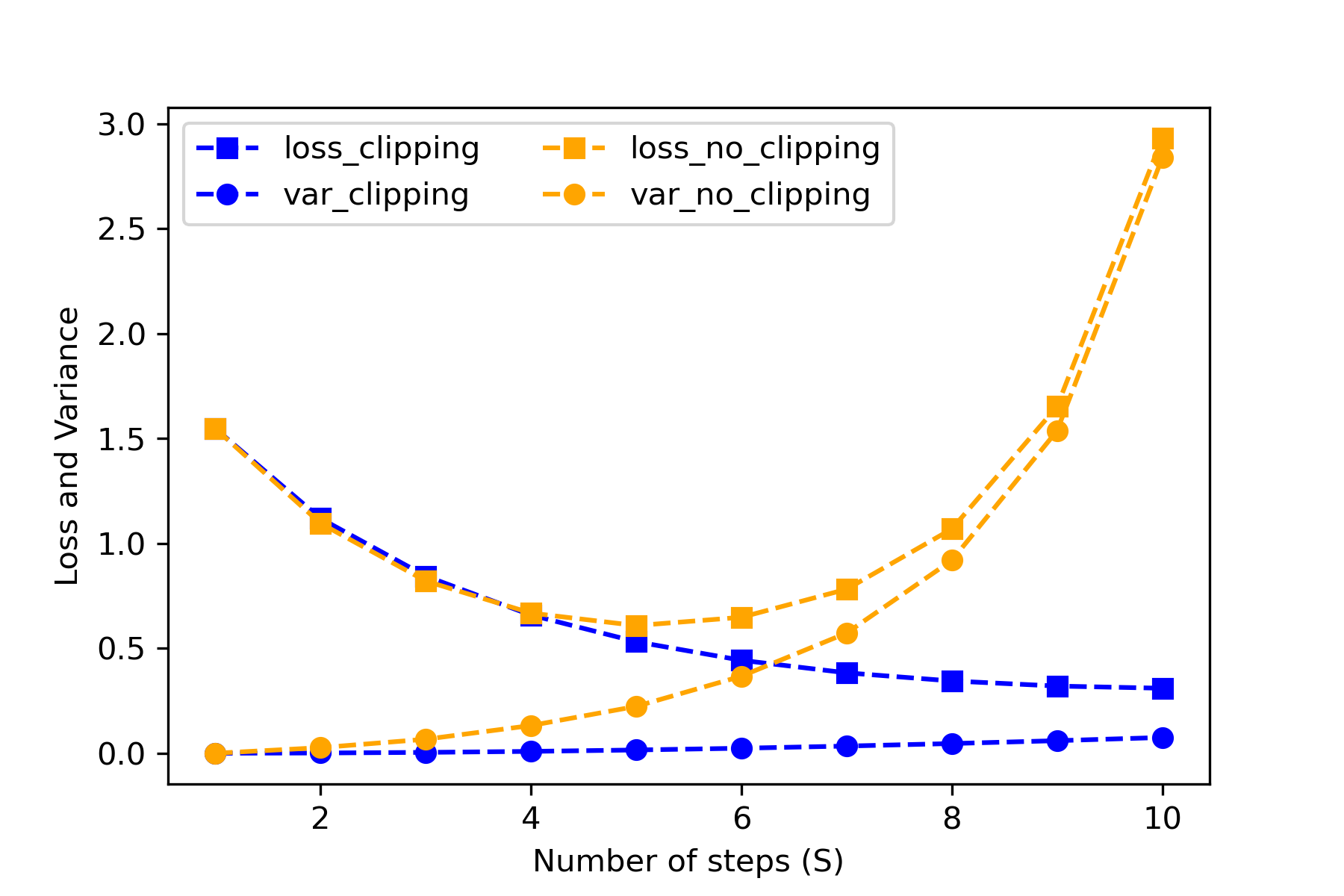}\label{fig:f1}}
  \hfill
  \subfloat[Varying the clipping constant $X$ on the Facebook graph]{\includegraphics[width=4cm]{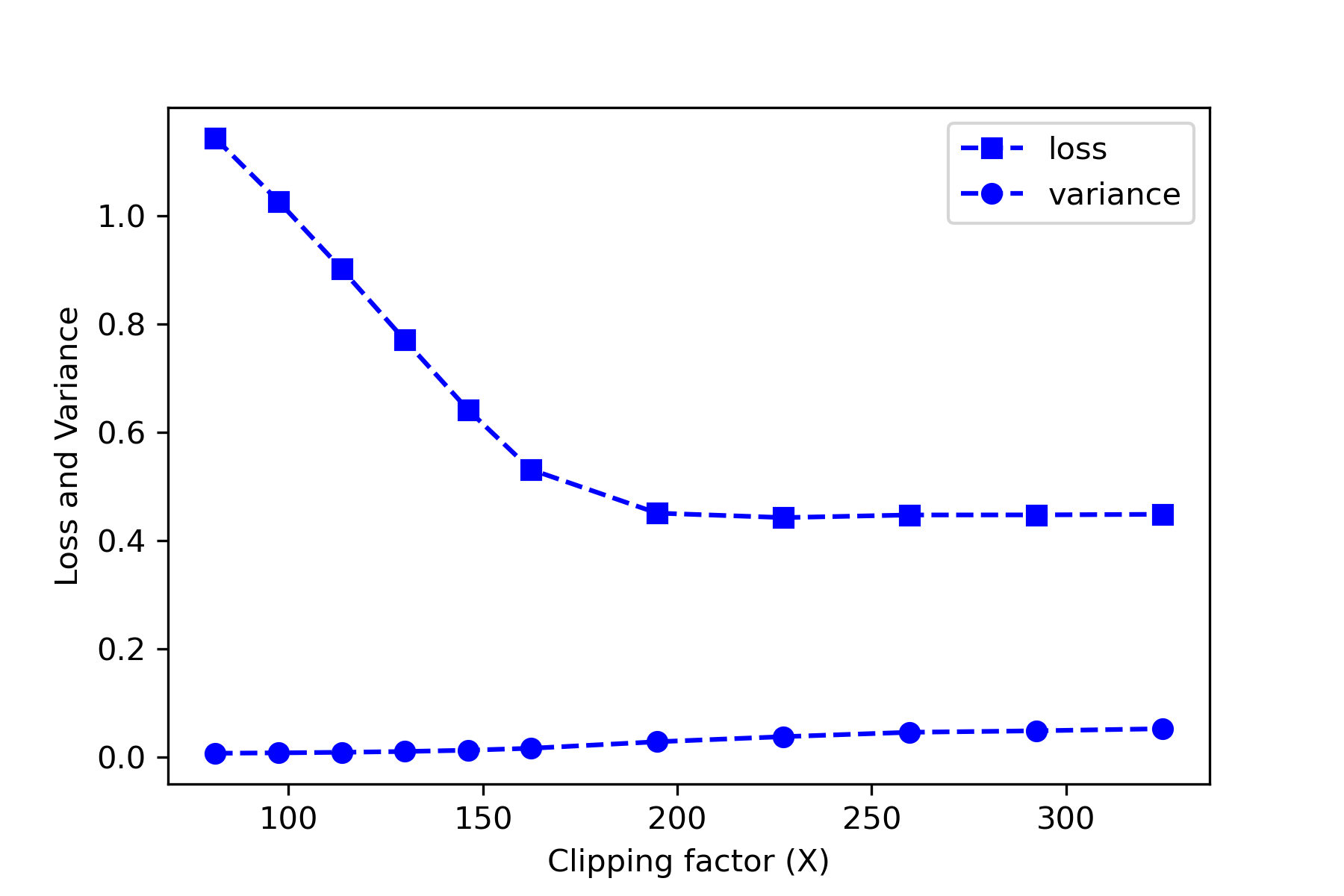}\label{fig:f2}}
  \hfill
  \subfloat[Varying the number of steps $S$ on the Wikipedia graph]{\includegraphics[width=4cm]{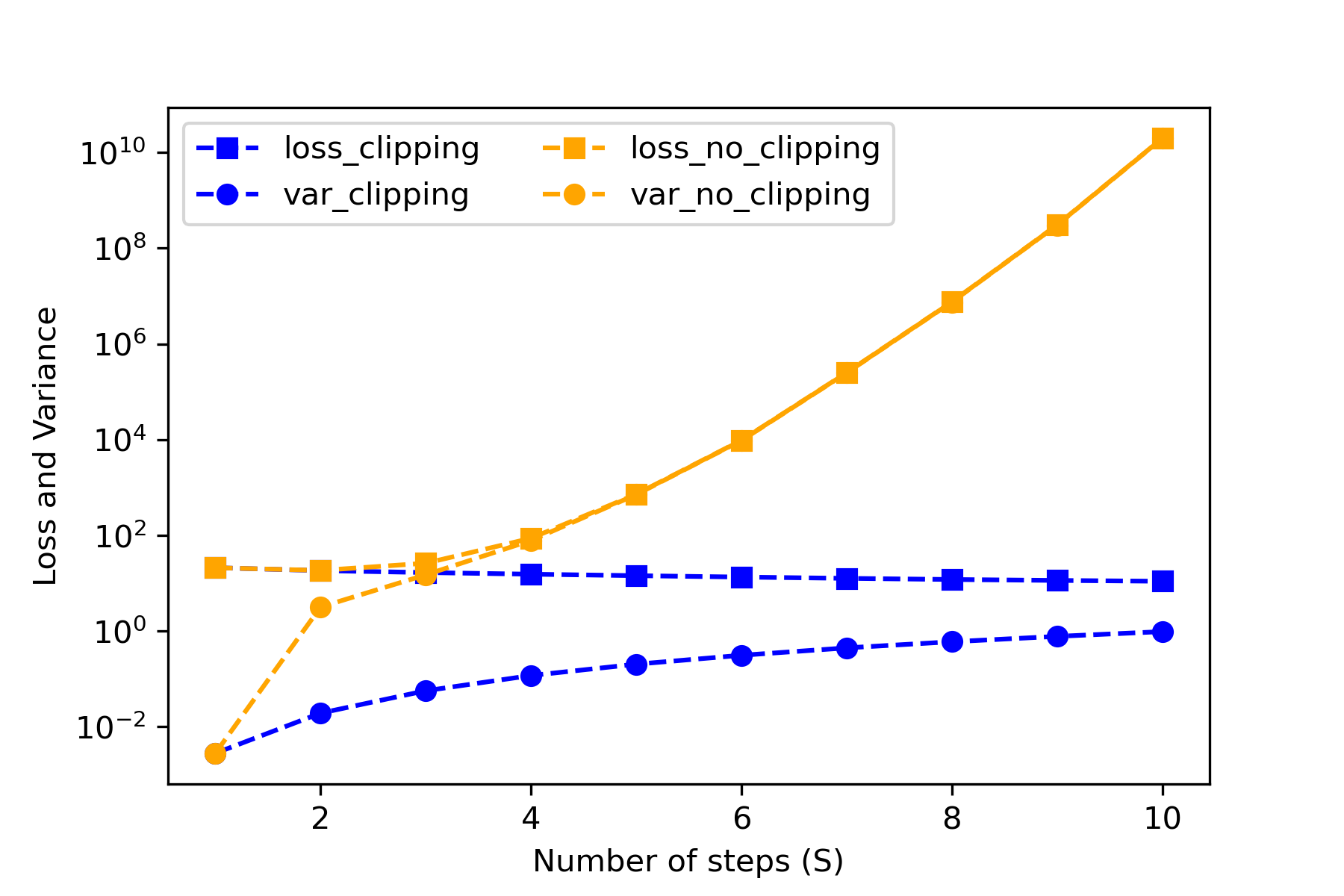}\label{fig:f3}}
  \hfill
  \subfloat[Varying the clipping constant $X$ on the Wikipedia graph]{\includegraphics[width=4cm]{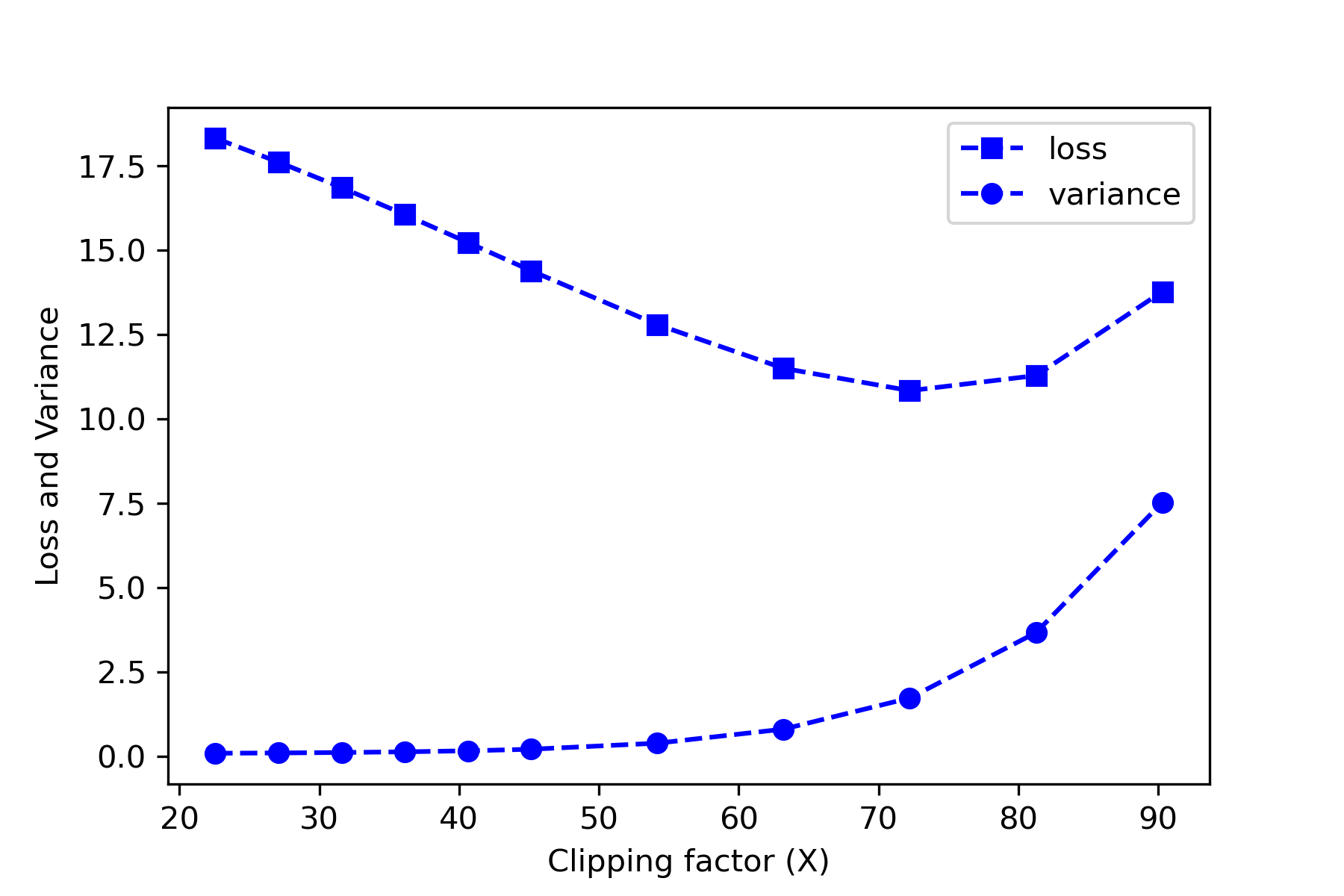}\label{fig:f4}}
  \vspace{-0.3cm}
  \captionsetup{justification = centering,font = small}
  \caption{Loss and variance of our Katz estimators (Algorithm~\ref{alg:LDPKatz}) \vspace{-0.7cm}}
\end{figure}

It is understood that the loss is the cumulative result of the variance and the square of the bias. Consequently, from the graphs, we can deduce the squared bias as the differential between the loss (represented by the squared line) and the variance (illustrated by the dotted line).

In \textbf{figure 1.a} and \textbf{1.c}, it is evident that with just one step, both algorithms exhibit a pronounced bias. This is because they only account for the number of paths of length 1 when estimating Katz centrality. Simultaneously, they display minimal variance due to the limited noise introduced. As the step count grows, meaning more paths are computed and more noise is integrated, we observe a decline in bias and a rise in variance across both algorithms. With increasing steps \(S\), the bias of the non-clipping algorithm (illustrated in orange) approaches zero since the Laplacian noise does not inherently introduce any bias. Yet, its variance surges sharply. Our theoretical evaluation from section 4.3 forecasts an exponential surge in variance when \(2 \alpha S H_{n}/\epsilon > 1\). This is corroborated in the experimental data, manifesting for \(S \geq 12\) in the Facebook graph and \(S \geq 4\) in the Wikipedia graph. As anticipated, the clipping mechanism (depicted in blue) effectively moderates the variance's escalation. While the bias does not reduce to zero as seen in the algorithm without clipping, it is evident that the bias remains sufficiently low. Consequently, the overall loss of our algorithm is minimal, making our publication impactful.

In \textbf{figure 1.b} and \textbf{1.d}, it's evident that the clipping factor helps lower the variance but at the expense of a higher bias. Specifically, when $X$ is small, there is a noticeable gap between the loss and the variance, which indicates a larger bias. In contrast, with larger values of $X$, the variance decreases. The algorithm without clipping is equivalent to having $X \rightarrow \infty$.

While the previous results provide valuable insights, in many practical scenarios, the primary concern is not the estimator's loss but its capability to identify the top $k$ nodes with the highest Katz centrality. Our next experiment focuses on this aspect. We ranked nodes based on the true Katz centrality values and compared them to rankings from our estimators. For specific values of $k$, we evaluated the percentage of top $k$ nodes, according to the real Katz centrality, that also appeared in the top $k$ nodes of each estimator. Figure 2 display the average detection rate of these top $k$ nodes, considering $k$ values of 10 and 100, along with confidence intervals. For \textbf{figure 2.a} and \textbf{2.c}, we applied a clipping factor $X$ equivalent to $E_{F}$ and $E_{W}$, respectively. Meanwhile, \textbf{figure 2.b} and \textbf{2.d} were based on 5 steps and clipping factors $X$ that varied between $0.5 E_{F}$ to $2 E_{F}$ and $0.5 E_{W}$ to $2 E_{W}$, respectively.

\begin{figure}
  \centering \vspace{-0.7cm}
  \subfloat[Varying the number of steps $S$ on the Facebook graph]{\includegraphics[width=4cm]{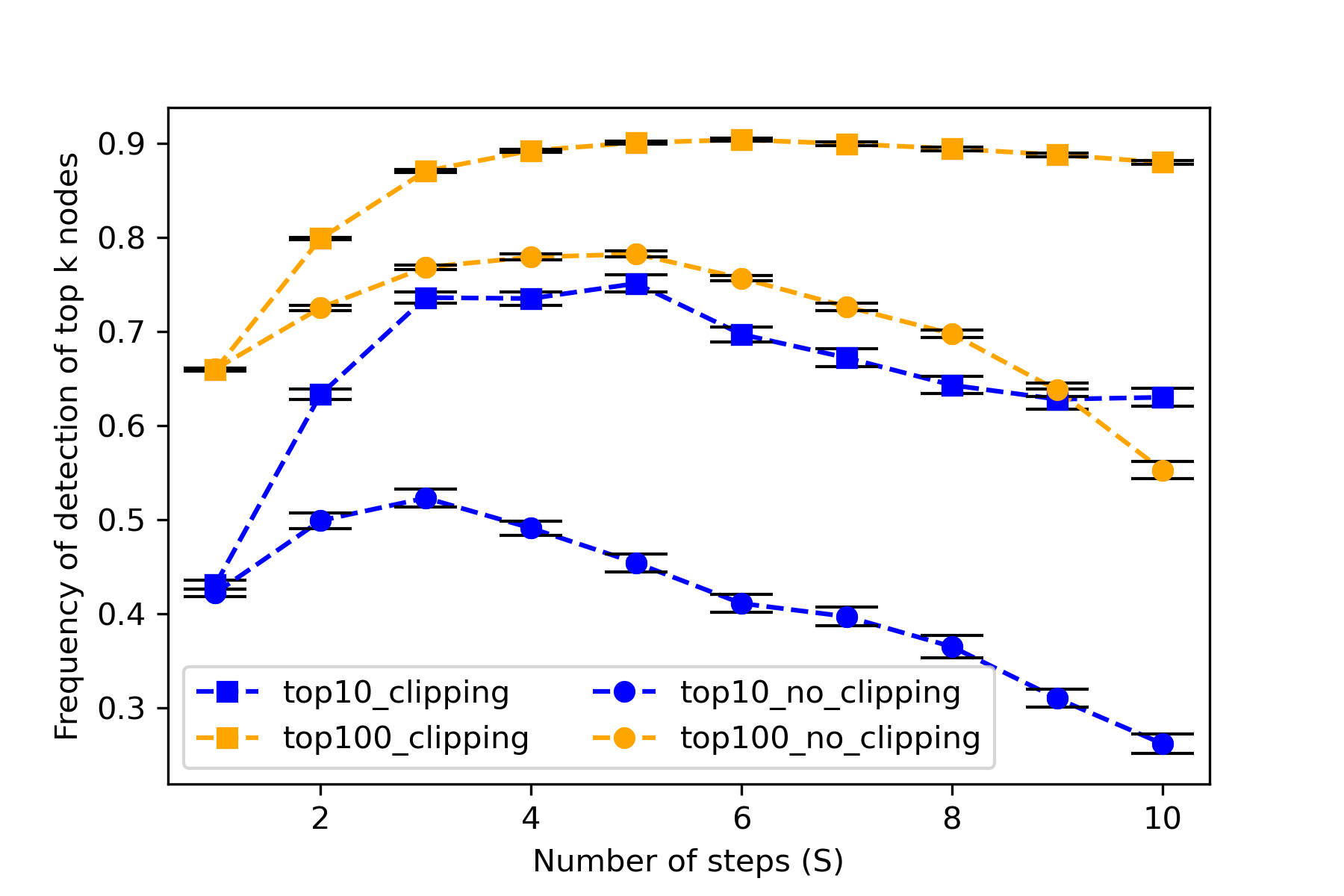}\label{fig:f5}}
  \hfill
  \subfloat[Varying the clipping constant $X$ on the Facebook graph]{\includegraphics[width=4cm]{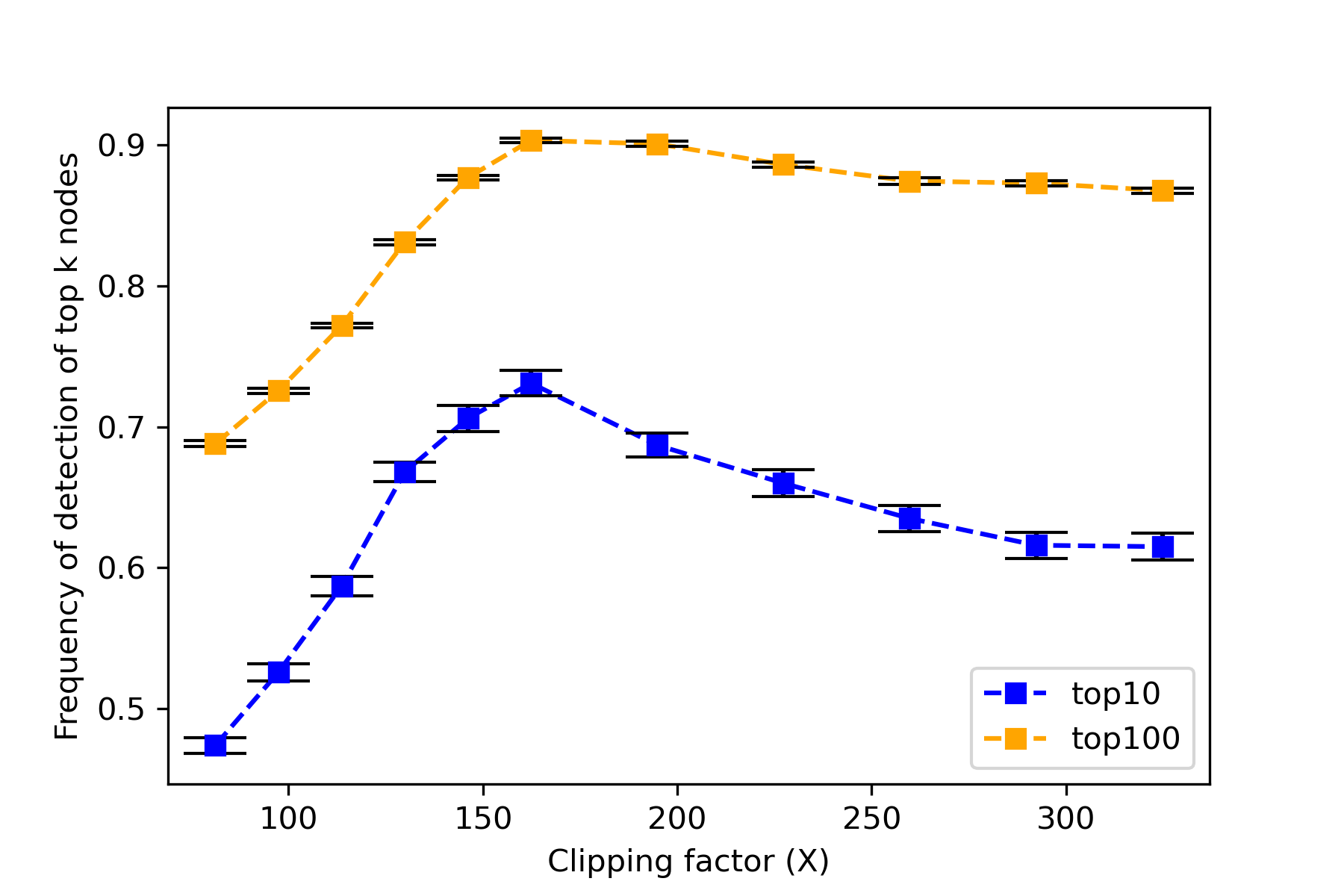}\label{fig:f6}}
  \hfill
  \subfloat[Varying the number of steps $S$ on the Wikipedia graph]{\includegraphics[width=4cm]{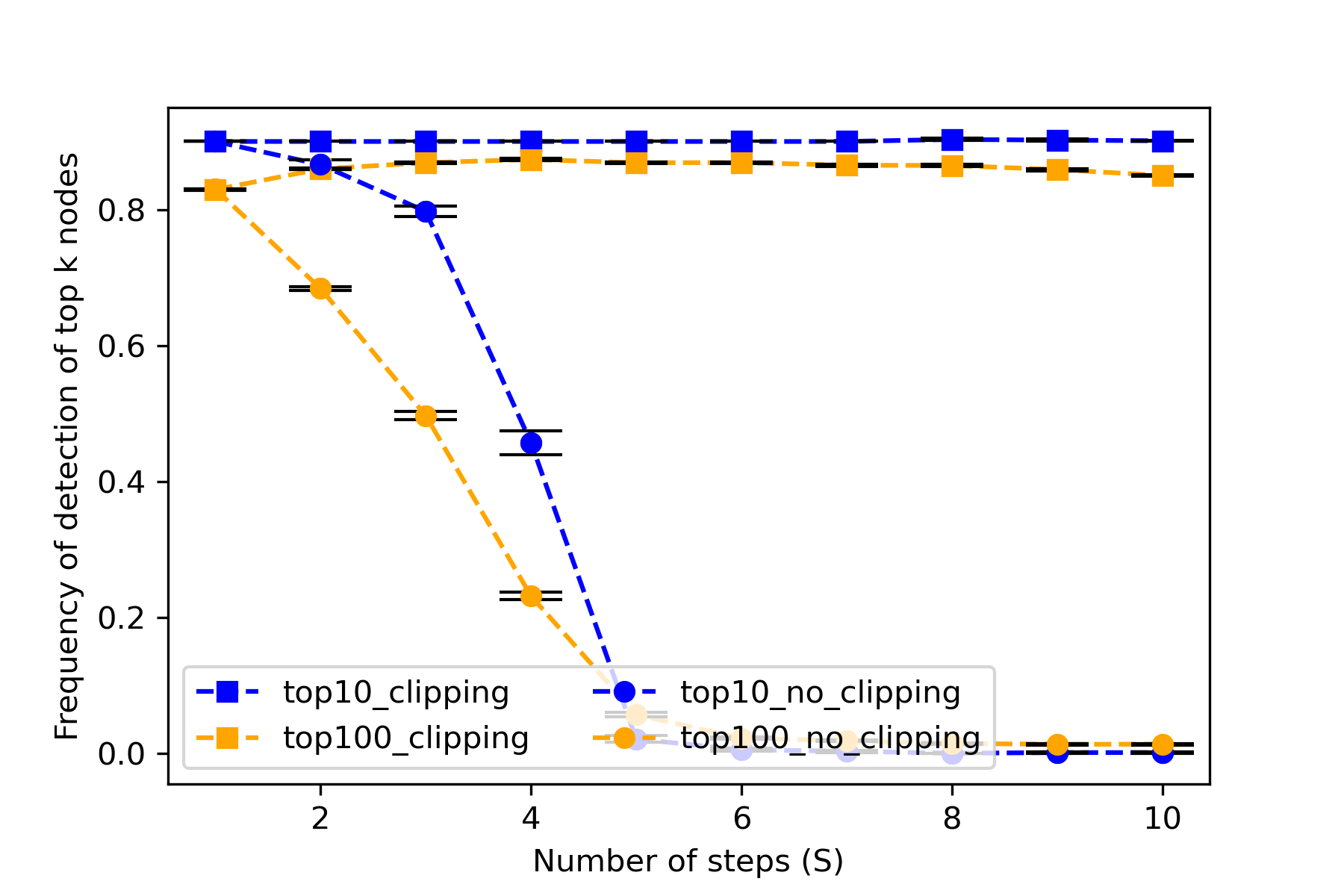}\label{fig:f7}}
  \hfill
  \subfloat[Varying the clipping constant $X$ on the Wikipedia graph]{\includegraphics[width=4cm]{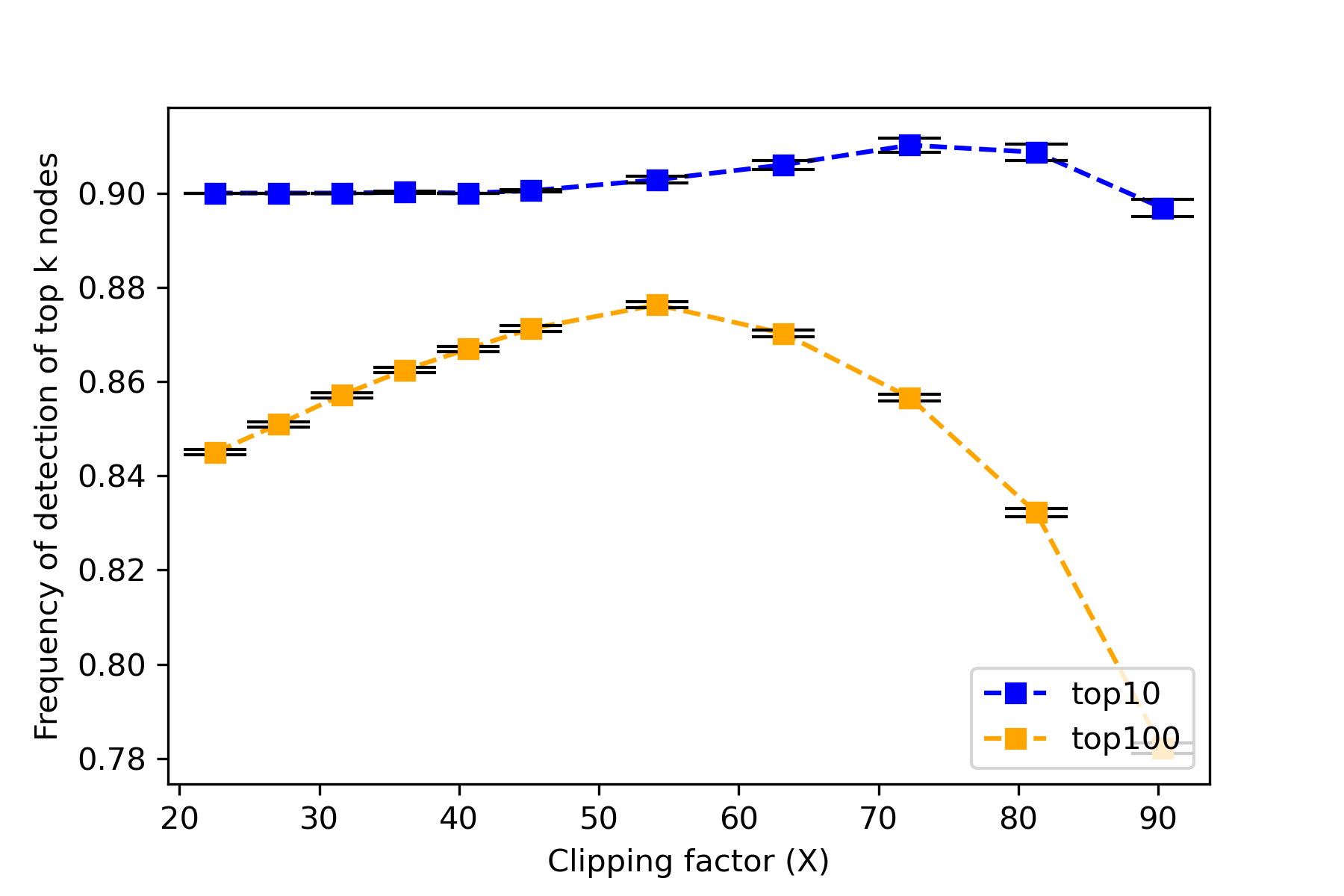}\label{fig:f8}}
  \captionsetup{justification = centering,font = small}
  \caption{Comparison of the top $k$ nodes determined by Katz centrality and those identified by our estimator} \vspace{-0.5cm}
\end{figure}

For the Facebook graph, our Katz centrality estimator successfully identifies about 90\% of the top 100 nodes and 73\% of the top 10 nodes. In the case of the Wikipedia graph, our estimator pinpoints 91\% of the top 10 nodes and close to 87\% of the top 100 nodes. We observe that the number of iterations, $S$, and the clipping constant, $X$, greatly influence recall rates. In our tests on both networks, setting $S$ to two or three and adjusting $X$ to match the maximum eigenvalues yielded the highest recall.

Finally, by matching the results from \textbf{figure 1} and \textbf{figure 2}, we notice that the best performance for detecting the top k nodes are not achieved for the smallest loss of the estimator but rather for smaller value of bias even at a cost of larger value of variance and overall loss. Indeed, for ranking nodes, a larger bias does not affect the performance as much as a larger variance since a larger bias means the Katz centrality estimation of every node have a similar negative error, possibly not changing the ranking, while adding noise almost certainly changes the ranking.

\subsection{Number of paths}

We also evaluated how well our algorithm estimates the number of paths. From \textbf{figure 3}, it is evident that the algorithm with clipping performs better than the one without. Notably, the non-clipping algorithm's variance shoots up exponentially when \(2 \alpha S H_{n} /\epsilon > 1\). This spike is observed when \( S \geq 12 \) for the Facebook graph and \( S \geq 4 \) for the Wikipedia graph.

\begin{figure}[htp] 
  \centering \vspace{-0.8cm}
  \subfloat[On the Facebook graph]{\includegraphics[width=4cm]{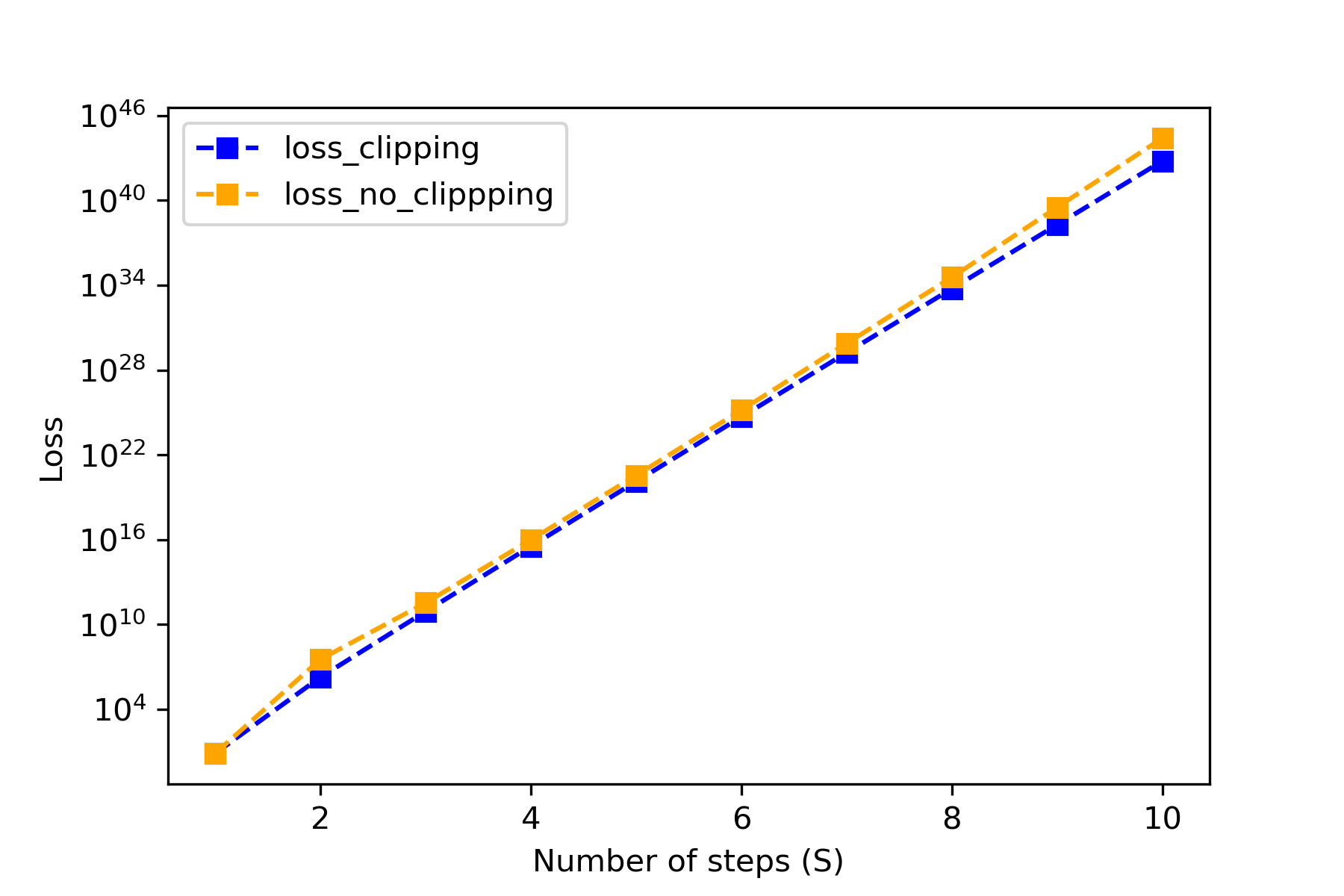}\label{fig:f9}}
  \hfill
  \subfloat[On the Wikipedia graph]{\includegraphics[width=4cm]{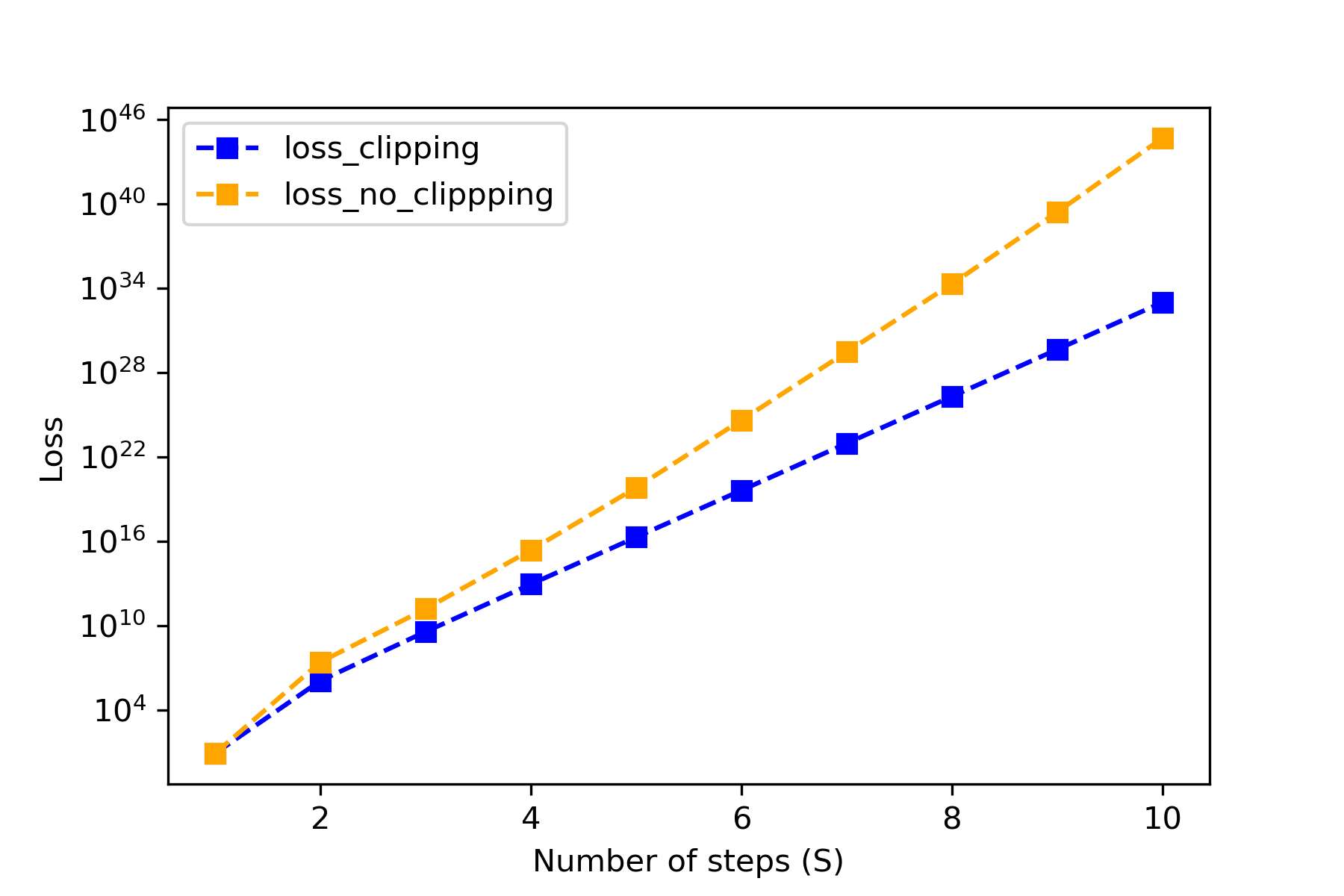}\label{fig:f10}}
  \captionsetup{justification = centering,font = small}
  \caption{Variation in loss for path estimators with increasing steps}
\end{figure}

\section{Conclusion}

 In this study, we developed an estimator to calculate both the number of paths and Katz centrality, leveraging multiple communication rounds and a clipping method. This approach maintains local differential privacy while effectively managing error. We provided an upper bound for the bias and variance associated with certain attenuation factor values, denoted as $\alpha$. Our findings also highlighted that, without the clipping method, the variance of our algorithm can escalate exponentially, even on the simplest of graphs. Our experiments further demonstrated that our algorithm performs well in ranking tasks, successfully identifying up to 90\% of the top $k$ nodes with the highest Katz centrality—a key metric in our research.

\bibliographystyle{ACM-Reference-Format}
\bibliography{main}

\end{document}